\documentstyle[aps,prl]{revtex}

\begin{document}
\draft
\date{October 16, 2001}
\title{Off--Diagonal 5D Metrics and Mass Hierarchies\\
with Anisotropies and Running Constants }
\author{Sergiu I. Vacaru}
\address{Department of Mathematics, University of Athens \\
15784\ Panepistimiopolis, Athens, Greece \\
and\\
Physics Department, California State University,\\
Fresno, CA 93740-8031, USA \\
{---} \\
E--mails: sergiu$_{-}$vacaru@yahoo.com,\\
 \ sergiuvacaru@venus.nipne.ro }
\maketitle

\begin{abstract}
The gravitational equations of the three dimensional (3D) brane world are
investigated for both off--diagonal and warped 5D metrics which can be
diagonalized with respect to some anholonomic frames when the gravitational
and matter fields dynamics are described by mixed sets of holonomic and
anholonomic variables. We construct two new classes of exact solutions of
Kaluza--Klein gravity which generalize the Randall--Sundrum metrics to
configurations with running on the 5th coordinate gravitational constant and
anisotropic dependencies of effective 4D constants on time and/or space
variables. We conclude that by introducing gauge fields as off--diagonal
components of 5D metrics, or by considering  anholonomic frames modelling
some anisotropies in extra dimension  spacetime, we induce anisotropic
tensions (gravitational  polarizations) and running of constants on the
branes. This way  we can generate the TeV scale as a hierarchically 
suppressed anisotropic mass scale and the Newtonian and general 
relativistic gravity are reproduced with adequate precisions but  with
corrections which depend anisotropically on some  coordinates.
\end{abstract}

\pacs{PACS numbers: Pacs 04.90.+e, 04.50.+h,}




Recent approaches to String/M--theory and particle physics are based on the
idea that our universe is realized as a three brane, modelling a four
dimensional, 4D, pseudo--Riemannian spacetime, embedded in the 5D anti--de
Sitter ($AdS_5$) bulk spacetime. In such models the extra dimension need not
be small (they could be even infinite) if a nontrivial warped geometric
configuration, being essential for solving the mass hierarchy problem and
localization of gravity, can ''bound'' the matter fields on a 3D subspace on
which we live at low energies, the gravity propagating, in general, in a
higher dimension spacetime (see Refs.: \cite{stringb} for string gravity
papers; \cite{arkani} for extra dimension particle fields and gravity
phenomenology with effective Plank scale; \cite{rs} for the simplest and
comprehensive models proposed by Randall and Sundrum; here we also point the
early works \cite{akama} in this line and cite \cite{shir} as some further
developments with supresymmetry, black hole solutions and cosmological
scenarios).

In higher di\-men\-sional gravi\-ty much at\-tention has been paid to
off--diago\-nal metrics beginning the Salam, Strathee and Perracci work \cite
{sal} which showed that including off--diagonal components in higher
dimensional metrics is equi\-valent to including $U(1),SU(2)$ and $SU(3)$
gauge fields. Recently, the off--diagonal metrics were considered in a new
fashion by applying the method of anholonomic frames with associated
nonlinear connections \cite{v} which allowed us to construct new classes of
solutions of Einstein's equations in three (3D), four (4D) and five (5D)
dimensions, with generic local anisotropy ({\it e.g.} static black hole and
cosmological solutions with ellipsoidal or torus symmetry, soliton--dilaton
2D and 3D configurations in 4D gravity, and wormhole and flux tubes with
anisotropic polarizations and/or running on the 5th coordinate constants
with different extensions to backgrounds of rotation ellipsoids, elliptic
cylinders, bipolar and torus symmetry and anisotropy.

The point of this paper is to argue that if the 5D gravitational
interactions are parametrized by off--diagonal metrics with a warped factor,
which could be related with an anholonomic higher dimensional gravitational
dynamics and/or with the fact that the gauge fields are included into a
Salam--Strathee --Peracci manner, the fundamental Plank scale $M_{4+d}$ in $%
4+d$ dimensions can be not only considerably smaller than the the effective
Plank scale, as in the usual locally isotropic Randall--Sundrum (in brief,
RS) scenarios, but the effective Plank constant is also anisotropically
polarized which could have profound consequences for elaboration of
gravitational experiments and for models of the very early universes.

We will give two examples with one additional dimension ($d=1$) when an
extra dimension gravitational anisotropic polarization on a space coordinate
is emphasized or, in the second case, a running of constants in time is
modelled. We will show that effective gravitational Plank scale is
determined by the higher-dimensional curvature and anholonomy of pentad
(funfbein, of frame basis) fields rather than the size of the extra
dimension. Such curvatures and anholonomies are not in conflict with the
local four-dimensional Poincare invariance.

We will present a higher dimensional scenario which provides a new RS like
approach generating anisotropic mass hierarchies. We consider that the 5D
metric is both not factorizable and off--diagonal when the four-dimensional
metric is multiplied by a ``warp'' factor which is a rapidly changing
function of an additional dimension and depend anisotropically on a space
direction and runs in the 5-th coordinate.

Let us consider a 5D pseudo--Riemannian spacetime provided with local
coordinates $u^\alpha =(x^i,y^a)=(x^1=x,x^2=f,x^3=y,y^4=s,y^5=p), $ where $%
\left( s,p\right) =\left( z,t\right) $ (Case I) or, inversely, $\left(
s,p\right) =\left( t,z\right) $ (Case II) -- or more compactly $u=(x,y)$ --
where the Greek indices are conventionally split into two subsets $x^i$ and $%
y^a$ labeled respectively by Latin indices of type $i,j,k,...=1,2,3$ and $%
a,b,...=4,5.$ The local coordinate bases, $\partial _\alpha =(\partial
_i,\partial _a),$ and their duals, $d^\alpha =\left( d^i,d^a\right) ,$ are
defined respectively as 
\begin{equation}
\partial _\alpha \equiv \frac \partial {du^\alpha }=(\partial _i=\frac %
\partial {dx^i},\partial _a=\frac \partial {dy^a}) \mbox{ and } d^\alpha
\equiv du^\alpha =(d^i=dx^i,d^a=dy^a).  \label{pdif}
\end{equation}

For the 5D (pseudo) Riemannian interval $dl^{2}=G_{\alpha \beta }du^{\alpha
}du^{\beta }$ we choose the metric coefficients $G_{\alpha \beta }$ (with
respect to the coordinate frame (\ref{pdif})) to be parametrized by a
off--diagonal matrix (ansatz) 
\begin{equation}
\left[ 
\begin{array}{ccccc}
g+w_{1}^{\ 2}h_{4}+n_{1}^{\ 2}h_{5} & w_{1}w_{2}h_{4}+n_{1}n_{2}h_{5} & 
w_{1}w_{3}h_{4}+n_{1}n_{3}h_{5} & w_{1}h_{4} & n_{1}h_{5} \\ 
w_{1}w_{2}h_{4}+n_{1}n_{2}h_{5} & 1+w_{2}^{\ 2}h_{4}+n_{2}^{\ 2}h_{5} & 
w_{2}w_{3}h_{4}+n_{2}n_{3}h_{5} & w_{2}h_{4} & n_{2}h_{5} \\ 
w_{1}w_{3}h_{4}+n_{1}n_{3}h_{5} & w_{3}w_{2}h_{4}+n_{2}n_{3}h_{5} & 
g+w_{3}^{\ 2}h_{4}+n_{3}^{\ 2}h_{5} & w_{3}h_{4} & n_{3}h_{5} \\ 
w_{1}h_{4} & w_{2}h_{4} & w_{3}h_{4} & h_{4} & 0 \\ 
n_{1}h_{5} & n_{2}h_{5} & n_{3}h_{5} & 0 & h_{5}
\end{array}
\right]   \label{ansatz0}
\end{equation}
where the coefficients are some necessary smoothly class functions of type: 
\begin{eqnarray*}
g &=&g(f,y)=a(f)\ b(y),h_{4}=h_{4}(f,y,s)=\eta _{4}(f,y)g(f,y)q_{4}(s), \\
h_{5} &=&h_{5}(f,y,s)=g(f,y)q_{5}(s),w_{i}=w_{i}(f,y,s),n_{i}=n_{i}(f,y,s).
\end{eqnarray*}

The metric (\ref{ansatz0}) can be equivalently rewritten in the form 
\begin{equation}
\delta l^{2}=g_{ij}\left( f,y\right) dx^{i}dx^{i}+h_{ab}\left( f,y,s\right)
\delta y^{a}\delta y^{b},  \label{dmetric}
\end{equation}
with diagonal coefficients 
\begin{equation}
g_{ij}=\left[ 
\begin{array}{lll}
g & 0 & 0 \\ 
0 & 1 & 0 \\ 
0 & 0 & g
\end{array}
\right] \mbox{ and }h_{ab}=\left[ 
\begin{array}{ll}
h_{4} & 0 \\ 
0 & h_{5}
\end{array}
\right]   \label{ansatzd}
\end{equation}
if instead the coordinate bases (\ref{pdif}) one introduce the anholonomic
frames (anisotropic bases) 
\begin{equation}
{\delta }_{\alpha }\equiv \frac{\delta }{du^{\alpha }}=(\delta _{i}=\partial
_{i}-N_{i}^{b}(u)\ \partial _{b},\partial _{a}=\frac{\partial }{dy^{a}})%
\mbox{ and }\delta ^{\alpha }\equiv \delta u^{\alpha }=(\delta
^{i}=dx^{i},\delta ^{a}=dy^{a}+N_{k}^{a}(u)\ dx^{k})  \label{ddif}
\end{equation}
where the $N$--coefficients are parametrized $N_{i}^{4}=w_{i}$ and $%
N_{i}^{5}=n_{i}.$

In this paper we consider a slice of $AdS_5$ provided with an anholnomic
frame structure (\ref{ddif}) satisfying the relations $\delta _\alpha \delta
_\beta -\delta _\beta \delta _\alpha =W_{\alpha \beta }^\gamma \delta
_\gamma ,$ with nontrivial anholonomy coefficients 
\[
W_{ij}^k =0,W_{aj}^k=0,W_{ia}^k=0,W_{ab}^k=,W_{ab}^c=0,  W_{ij}^a =\delta
_iN_j^a-\delta _jN_i^a,W_{bj}^a=-\partial _bN_j^a,W_{ia}^b=\partial _aN_j^b. 
\]

We assume there exists a solution of 5D Einstein equations with 3D brane
configuration that effectively respects the local 4D Poincare invariance
with respect to anholonomic frames (\ref{ddif}) and that the metric ansatz (%
\ref{ansatz0}) (equivalently, (\ref{ansatzd})) transforms into the usual RS
solutions 
\begin{equation}
ds^2=e^{-2k|f|}\eta _{\underline{\mu }\underline{\nu }}dx^{\underline{\mu }%
}dx^{\underline{\nu }}+df^2  \label{solrs}
\end{equation}
for the data:\ $a(f) = e^{-2k|f|},\ k=const,\ b(y)=1, \eta _4(f,y) = 1,
q_4(s)=q_5(s)=1, w_i = 0,n_i=0, $ where $\eta _{\underline{\mu }\underline{%
\nu }}$ and $x^{\underline{\mu }}$ are correspondingly the diagonal metric
and Cartezian coordinates in 4D Minkowski spacetime and the
extra-dimensional coordinate $f$ is to be identified $f=r_c\phi ,$ ($%
r_c=const$ is the compactification radius, $0\leq f\leq \pi r_c$) like in
the first work \cite{rs} (or 'f'' is just the coordinate 'y' in the second
work \cite{rs}).

The set-up for our model is a single 3D brane with positive tension,
subjected to some anholonomic constraints, embedded in a 5D bulk spacetime
provided with a off--diagonal metric (\ref{ansatz0}). In order to carefully
quantize the system, and treat the non-normalizable modes which will appear
in the Kaluza-Klein reduction, it is useful to work with respect to
anholonomic frames were the metric is diagonalized by corresponding
anholonomic transforms and is necessary to work in a finite volume by
introducing another brane at a distance $\pi r_c$ from the brane of
interest, and taking the branes to be the boundaries of a finite 5th
dimension. We can remove the second brane from the physical set-up by taking
the second brane to infinity.

The action for our anholonomic funfbein (pentadic) system is 
\begin{eqnarray}
S &=&S_{gravity}+S_{brane}+S_{brane^{\prime }}  \label{anhbr} \\
S_{gravity} &=&\int \delta ^{4}x\int \delta f\sqrt{-G}\{-\Lambda
(f)+2M^{3}R\},\ S_{brane}=\int \delta ^{4}x\sqrt{-g_{brane}}\{V_{brane}+%
{\cal L}_{brane}\},  \nonumber
\end{eqnarray}
where $R$ is the 5D Ricci scalar made out of the 5D metric, $G_{\alpha \beta
}$, and $\Lambda $ and $V_{brane}$ are cosmological terms in the bulk  and
boundary respectively. We write down $\delta ^{4}x$ and $\delta f,$ instead
of usual differentials $d^{4}x$ and $df,$ in order to emphasize that the
variational calculus should be performed by using N--elongated partial
derivatives and differentials (\ref{ddif}). The coupling to the branes and
their fields and the related orbifold boundary conditions for vanishing
N--coefficients are described in Refs. \cite{rs} and \cite{me1}.

The Einstein equations, $R_{\beta }^{\alpha }-\frac{1}{2}\delta _{\beta
}^{\alpha }R=\Upsilon _{\beta }^{\alpha },$ for a diagonal
energy--moment\-um tensor $\Upsilon _{\alpha }^{\beta }=\left[ \Upsilon
_{1},\Upsilon _{2},\Upsilon _{3},\Upsilon _{4},\Upsilon _{5}\right] $ and
following from the action (\ref{anhbr}) and for the ansatz (\ref{ansatz0})
(equivalently, (\ref{ansatzd})) with $g=a(f)b(y)$ transform into 
\begin{eqnarray}
\frac{1}{a}\left[ a_{1}^{\prime \prime }-\frac{(a^{\prime })^{2}}{2a}\right]
+\frac{\beta }{h_{4}h_{5}} &=&2\Upsilon _{1},\ \frac{(a^{\prime })^{2}}{2a}+%
\frac{P(y)}{a}+\frac{\beta }{h_{4}h_{5}}=2\Upsilon _{2}(f),  \label{einst} \\
\frac{1}{a}\left[ a^{\prime \prime }-\frac{(a^{\prime })^{2}}{2a}\right] +%
\frac{P(y)}{a} &=&2\Upsilon _{4},\ w_{i}\beta +\alpha _{i}=0,\ n_{i}^{\ast
\ast }+\gamma n_{i}^{\ast }=0,  \nonumber
\end{eqnarray}
where 
\begin{eqnarray}
\alpha _{1} &=&\ {h_{5}^{\ast }}^{^{\bullet }}-\frac{{h_{5}^{\ast }}}{2}%
\left( \frac{h_{4}^{\bullet }}{h_{4}}+\frac{h_{5}^{\bullet }}{h_{5}}\right)
\ \alpha _{2}={h_{5}^{\ast }}^{^{\prime }}-\frac{{h_{5}^{\ast }}}{2}\left( 
\frac{h_{4}^{^{\prime }}}{h_{4}}+\frac{h_{5}^{^{\prime }}}{h_{5}}\right)
,\alpha _{3}={h_{5}^{\ast }}^{\#}-\frac{{h_{5}^{\ast }}}{2}\left( \frac{%
h_{4}^{\#}}{h_{4}}+\frac{h_{5}^{\#}}{h_{5}}\right) ,  \nonumber \\
\beta  &=&h_{5}^{\ast \ast }-\frac{(h_{5}^{\ast })^{2}}{2h_{5}}-\frac{%
h_{5}^{\ast }h_{4}^{\ast }}{2h_{4}},P=\frac{1}{b^{2}}\left[ b^{\#\#}-\frac{%
(b^{\#})^{2}}{b}\right] ,\ \gamma =\frac{3}{2}\frac{h_{5}}{h_{5}}^{\ast }-%
\frac{h_{4}}{h_{4}}^{\ast },  \label{gamma}
\end{eqnarray}
the partial derivatives are denoted: $h^{\bullet }=\partial h/\partial
x^{1},h^{\prime }=\partial h/\partial x^{2},h^{\#}=\partial h/\partial x^{3},
$ $h^{\ast }=\partial h/\partial s.$

Our aim is to construct a metric 
\begin{equation}
\delta s^{2}=g\left( f,y\right) [dx^{2}+dy^{2}+\eta _{4}\left( f,y\right)
\delta s^{2}+q_{5}(s)\delta p^{2}]+df^{2},  \label{sol1b}
\end{equation}
with the anholonomic frame components defined by 'elongation' of
differentials, $\delta s=ds+w_{2}df+w_{3}dy,$ $\delta
p=dp+n_{1}dx+n_{2}df+n_{3}dy,$ and the ''warp'' factor being written in a
form similar to the RS solution 
\begin{equation}
g\left( f,y\right) =a(f)b(y)=\exp [-2k_{f}|f|-2k_{y}|y|],  \label{warpan}
\end{equation}
which defines anisotropic RS like solutions of 5D Einstein equations with
variation on the 5th coordinate cosmological constant in the bulk and
possible variations of induced on the brane cosmological constants.

By straightforward calculations we can verify that a class of exact
solutions of the system of equations (\ref{einst}) for $P(y)=0\,$ (see (\ref
{gamma})) :
\[
h_{4}=g(f,y),h_{5}=g(f,y)\rho ^{2}(f,y,s),
\]
were 
\begin{eqnarray*}
\rho (f,y,s) &=&|\cos \tau _{+}\left( f,y\right) |,\ \tau _{+}=\sqrt{\left(
\Upsilon _{4}-\Upsilon _{2}\right) g(f,y)},\Upsilon _{4}>\Upsilon _{2}; \\
&=&\exp [-\tau _{-}\left( f,y\right) s],\tau _{-}=\sqrt{\left( \Upsilon
_{2}-\Upsilon _{4}\right) g(f,y)},\Upsilon _{4}<\Upsilon _{2}; \\
&=&|c_{1}(f,y)+sc_{2}(f,y)|^{2},\Upsilon _{4}=\Upsilon _{2},
\end{eqnarray*}
\ and 
\begin{eqnarray*}
w_{i} &=&-\partial _{i}(\ln |\rho ^{\ast }|)/(\ln |\rho ^{\ast }|)^{\ast },
\\
n_{i} &=&n_{i[0]}(f,y)+n_{i[1]}(f,y)\int \exp [-3\rho ]ds,
\end{eqnarray*}
with functions  $c_{1,2}(f,y)$ and $n_{i[0,1]}(f,y)$  to be stated by some
boundary conditions.  We emphasize that the constants $k_{f}$ and $k_{y}$
have to be defined from some experimental data. 

The solution (\ref{sol1b}) transforms into the usual RS solution (\ref{solrs}%
) if $k_{y}=0,n_{i[0,1]}(f,y)=0,\Lambda =\Lambda _{0}=const$ and $\Upsilon
_{2}\rightarrow \Upsilon _{2[0]}=-{\frac{\Lambda _{0}}{4M^{3}};}\ \Upsilon
_{1},\Upsilon _{3},\Upsilon _{4},\Upsilon _{5}\rightarrow \Upsilon _{\lbrack
0]}=\frac{V_{brane}}{4M^{3}}\delta (f)+\frac{V_{brane^{\prime }}}{4M^{3}}%
\delta (f-\pi r_{c}),$ which holds only when the boundary and bulk
cosmological terms are related by formulas $V_{brane}=-V_{brane^{\prime
}}=24M^{3}k_{f},~~\Lambda _{0}=-24M^{3}k_{f}^{2};\ $ we use values with the
index $[0]$ in order to emphasize that they belong to the usual (holonomic)
RS solutions. In the anholonomic case with ''variation of constants'' we
shall not impose such relations.

We note that using the metric (\ref{sol1b}) with anisotropic warp factor (%
\ref{warpan}) it is easy to identify the massless gravitational fluctuations
about our classical solutions like in the usual RS cases but performing (in
this work) all computations with respect to anholonomic frames. All
off--diagonal fluctuations of the anholonomic diagonal metric are massive
and excluded from the low-energy effective theory.

We see that the physical mass scales are set by an anisotropic
symmetry--breaking scale, $v(y)\equiv e^{-k_{y}|y|}e^{-k_{f}r_{c}\pi }v_{0}.$%
\ This result the conclusion: any mass parameter $m_{0}$ on the visible
3-brane in the fundamental higher-dimensional theory with Salam--Strathee
--Peracci gauge interactions and/or effective anholonomic frames will
correspond to an anisotropic dependence on coordinate $y$ of the physical
mass $m(y)\equiv e^{-k_{y}|y|}e^{-kr_{c}\pi }m_{0}$ when measured with the
metric $\overline{g}_{\mu \nu }$ that appears in the effective Einstein
action, since all operators get rescaled according to their four-dimensional
conformal weight. If $e^{kr_{c}\pi }$ is of order $10^{15}$, this mechanism
can produces TeV physical mass scales from fundamental mass parameters not
far from the Planck scale, $10^{19}$ GeV. Because this geometric factor is
an exponential, we clearly do not require very large hierarchies among the
fundamental parameters, $v_{0},k,M,$ and $\mu _{c}\equiv 1/r_{c}$; in fact,
we only require $kr_{c}\approx 50$. These conclusions were made in Refs. 
\cite{rs} with respect to diagonal (isotropic) metrics. But the physical
consequences could radically change if the off--diagonal metrics with
effective anholonomic frames and gauge fields are considered. In this case
we have additional dependencies on variable $y$ which make the fundamental
spacetime geometry to be locally anisotropic, polarized via dependencies
both on coordinate $y$  receptivitity $k_{y}.$ We emphasize that our $y$
coordinate is not that from \cite{rs}.

The phenomenological implications of these anisotropic scenarios for future
collider searches could be very distinctive: the geometry of experiments
will play a very important role. In such anisotropic models we also have a
roughly weak scale splitting with a relatively small number of excitations
which can be kinematically accessible at accelerators.

We also reconsider in an anisotropic fashion the derivation of the 4D
effective Planck scale $M_{Pl}$ given in Ref. \cite{rs}. The 4D graviton
zero mode follows from the solution, Eq. (\ref{sol1b}), by replacing the
Minkowski metric by a effective 4D metric $\overline{g}_{\mu \nu }$ which it
is described by an effective action following from substitution into Eq. (%
\ref{anhbr}), 
\begin{equation}
S_{eff}\supset \int \delta ^{4}x\int_{0}^{\pi
r_{c}}df~2M^{3}r_{c}e^{-2k_{f}|f|}e^{-2k_{y}|y|}\sqrt{\overline{g}}~%
\overline{R},  \label{effaction}
\end{equation}
where $\overline{R}$ denotes the four-dimensional Ricci scalar made out of $%
\overline{g}_{\mu \nu }(x)$, in contrast to the five-dimensional Ricci
scalar, $R$, made out of $G_{MN}(x,f)$. We use the symbol $\delta ^{4}x$ in (%
\ref{anhbr}) in order to emphasize that our integration is adapted to the
anholonomic strucute stated by the differentials (\ref{ddif}). We also can
explicitly perform the $f$ integral in (\ref{effaction}) to obtain a purely
4D action and to derive 
\begin{equation}
M_{Pl}^{2}=2M^{3}\int_{0}^{\pi r_{c}}dfe^{-2k_{f}|f|}=\frac{M^{3}}{k}%
e^{-2k_{y}|y|}[1-e^{-2k_{f}r_{c}\pi }].  \label{effplanck}
\end{equation}
We see that there is a well-defined value for $M_{Pl}$, even in the $%
r_{c}\rightarrow \infty $ limit, but which may have an anisotropic
dependence on one of the 4D coordinates, in the stated parametrizations
denoted by $y$. Nevertheless, we can get a sensible effective anisotropic 4D
theory, with the usual Newtonian force law, even in the infinite radius
limit, in contrast to the product--space expectation that $%
M_{Pl}^{2}=M^{3}r_{c}\pi $.

In consequence of (\ref{effplanck}), the gravitational potential behaves
anisotropically as 
\[
V(r)=G_{N}{\frac{m_{1}m_{2}}{r}}\left( 1+\frac{e^{-2k_{y}|y|}}{r^{2}k_{f}^{2}%
}\right) 
\]
i.e. our models produce effective 4D theories of gravity with local
anisotropy. The leading term due to the bound state mode is the usual
Newtonian potential; the Kaluza Klein anholonomic modes generate an
extremely anisotropically suppressed correction term, for $k_{f}$ taking the
expected value of order the fundamental Planck scale and $r$ of the size
tested with gravity.

Let us conclude the paper: It is known that we can consistently exist with
an infinite 5th dimension, without violating known tests of gravity \cite{rs}%
. The scenarios consist of two or a single 3-brane, (a piece of) $AdS_5$ in
the bulk, and an appropriately tuned tension on the brane. But if we
consider off--diagonal 5D metrics like in Ref. \cite{sal}, which was used
for including of $U(1),SU(2)$ and $SU(3)$ gauge fields, or, in a different
but similar fashion, for construction of generic anisotropic, partially
anholonomic, solutions (like static black holes with ellipsoidal horizons,
static black tori and anisotropic wormholes) in Einstein and extra dimension
gravity, \cite{v} the RS theories become substantially locally anisotropic.
One obtains variations of constants on the 5th coordinate and possible
anisotropic oscillations in time (in the first our model), or on space
coordinate (in the second our model). Here it should be emphasized that the
anisotropic oscillations (in time or in a space coordinate) are defined by
the constant component of the cosmological constant (which in our model can
generally run on the 5th coordinate). This sure is related to the the
cosmological constant problem which in this work is taken as a given one,
with an approximation of linear dependence on the 5th coordinate, and not
solved. In the other hand a new, anisotropic, solution to the hierarchy
problem is supposed to be subjected to experimental verification.

Finally, we note that many interesting questions remain to be investigated.
Having constructed another, anisotropic, valid alternative to conventional
4D gravity, it is important to analyze the astrophysical and cosmological
implications. These anisotropic scenarios might even provide a new
perspective for solving unsolved issues in string/M-theory, quantum gravity
and cosmology. \vskip 4pt

~~ The author thanks D. Singleton, E. Gaburov and D. Gontsa for
collaboration and discussing of results. The author is grateful to P.
Stavrinos for hospitality and support. The work is partially supported by
''The 2000--2001 California State University Legislative Award''.



\end{document}